          [2001/04/25 1.1 (PWD)]
\documentclass[a4paper,twocolumn]{esapub2005} 
\usepackage{times}
\usepackage{natbib}

\usepackage{pstcol}
\usepackage[latin1]{inputenc}
\usepackage[T1]{fontenc}
\usepackage{pstcol}
\usepackage{amssymb}
\usepackage{amsmath}
\usepackage{txfonts}
\usepackage{graphicx}
\usepackage{xspace}
\usepackage{natbib}
\usepackage{amsmath}
\usepackage{txfonts}
\usepackage{url}

\newcommand{\cyg}{\mbox{Cyg~X-1}\xspace}
\newcommand{\inte}{\textsl{INTEGRAL}\xspace}
\newcommand{\xte}{\textsl{RXTE}\xspace}
\newcommand{\xmm}{\textsl{XMM-Newton}\xspace}
\newcommand{\kev}{\ensuremath{\text{ke\kern -0.09em V}}\xspace}
\newcommand{\pca}{\textsl{PCA}\xspace}
\newcommand{\hexte}{\textsl{HEXTE}\xspace}
\newcommand{\asm}{\textsl{ASM}\xspace}
\newcommand{\ibis}{\textsl{IBIS (ISGRI)}\xspace}
\newcommand{\spi}{\textsl{SPI}\xspace}
\newcommand{\jemx}{\textsl{JEM-X}\xspace}
\newcommand{\xspec}{\textsl{XSPEC}\xspace}
\newcommand{\eqpair}{\texttt{eqpair}\xspace}
\newcommand{\msun}{\ensuremath{\text{M}_{\odot}}}

\title{The broadband spectrum of Cygnus X-1}

\author[1]{S.~Fritz}
\author[2]{J.~Wilms}
\author[3]{K.~Pottschmidt}
\author[4]{M. A.~Nowak}
\author[1]{E.~Kendziorra}
\author[5]{M. G.~Kirsch}
\author[1,6]{I.~Kreykenbohm}
\author[1]{A.~Santangelo}
\affil[1]{Institut f\"ur Astronomie und Astrophysik, Sand 1, 72076 T\"ubingen, Germany} 
\affil[2]{Dr.\ Remeis Sternwarte, Astronomisches Institut der Universit\"at
Erlangen-N\"urnberg, Sternwartstr.~7, 96049 Bamberg, Germany}
\affil[3]{Center for Astrophysics and Space Sciences, University of
  California, San Diego, La Jolla, CA 92093-0424, USA}
\affil[4]{MIT-CXC, NE80-6077, 77 Massachusetts Ave., Cambridge, MA 02139, USA}
\affil[5]{European Space Astronomy Centre (ESA), Madrid, Spain}
\affil[6]{INTEGRAL Science Data Centre, 16 Ch. d'Ecogia, 1290 Versoix, Switzerland}

\begin{document}


\maketitle

\begin{abstract}
The Black Hole (BH) binary Cygnus X-1 has been observed
simultaneously by \inte, \xte, and \xmm for four times in November and
December 2004, when \cyg became first observable with \xmm. During these
observations the source was found in one of its transitional states between
the hard state and the soft state. We obtained a high signal to noise spectrum
of \cyg from 3\,keV to 1\,MeV which allows us to put constraints on the
nature of the Comptonizing plasma by modeling the continuum with
Comptonization models as \eqpair \citep{coppi92}. Using \xmm we were also able to confirm the presence of
a relativistically broadened Fe K$\alpha$ line.
\end{abstract}

\section{Introduction}
Being one of the brightest sources in the X-ray sky, \cyg has become also one
of the best studied Galactic BHs. The system consists of the O9.7I star
HDE~226868 of 40\,\msun \citep{ziolkowski05} and a compact object with a mass
of about 10\,\msun. Most of the time (90\% up to MJD~51300, 75\% from
then on \citep{wilms06a}) the source can be found in its hard state, where the X-ray
spectrum is characterized by a power law with photon index $\Gamma \approx
1.7$ and an exponential cutoff at $\approx 150\,\kev$
\citep[][and references therein]{gierlinski97}. Further spectral
characteristics are reflection features from the accretion disk and a
relativistically broadened Fe K$\alpha$ emission line at about 6.4\,\kev. The
other canonical state in which \cyg can be found is the soft state during
which the spectrum is dominated by a soft component peaking at $\sim 1\,\kev$
followed by a power law tail with index $\Gamma \approx 2 - 3$
\citep{dolan77,ogawara82}. Between these two extremes
\cyg can also be found in the so called ``intermediate states'' which exhibit
properties of both canonical states \citep{belloni05}. 

\begin{figure}
\centering
\includegraphics[width=\linewidth]{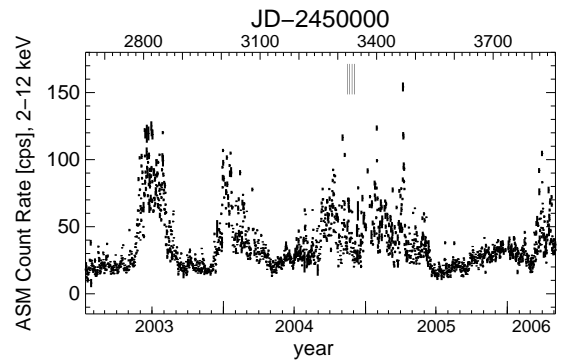}
\caption{\xte \asm light curve of \cyg. The vertical lines mark the times of our
  observations.\label{fig:lc}}
\end{figure}

\cyg was observed in 2004 with \inte, \xte, and \xmm simultaneously
on November 14/15 (hereafter called obs1), 20/21 (obs2), 26/27 (obs3) and on
December 2/3 (obs4). Fig.\ref{fig:lc} shows the
\xte/\asm lightcurve of \cyg with the four observations indicated by vertical
bars. The log of observations is given in
Table~\ref{tab:log}. The observations took place during one of the transitional states of
\cyg  which are very interesting for a
further study due to the fact that they are characterized by radio flaring, the
presence of a relativistically broadened iron line as well as a complex X-ray
timing behavior \citep[e.g.][]{cui97,miller02,wilms06a,malzac06,cadollebel06}. 

\begin{table}
\centering
\caption{Observing log for the three instruments used. Times are given in ksec.\label{tab:log}}
\begin{tabular}{lccc}
\hline
 Date          & \textsl{XMM}   & \textsl{RXTE}  & \textsl{INTEGRAL} \\
\hline
14/15 Nov 04 (obs1) & 17.6 & 12.0 &  58.0   \\
20/21 Nov 04 (obs2) & 17.7 & 16.7 &  80.0   \\
26/27 Nov 04 (obs3) & 20.0 & 26.0 &  79.0   \\
02/03 Dec 04 (obs4) & 10.0 & 10.0 &  80.0   \\
\hline
\end{tabular}
\end{table}

In this contribution we present some results of our ongoing analysis. For \xte
we used data from the \pca and \hexte, covering an energy range from 3 to
120\,{\ensuremath{\text{ke\kern -0.09em V}}}. The data extraction was done
using HEASOFT 5.3.1. The \inte spectra comprise information of the three
instruments \jemx, \ibis, and \spi, including energies up to
1\,{\ensuremath{\text{Me\kern -0.09em V}}} and were reduced using \inte
OSA~5.1. The data reduction followed the standard procedures as described in
the cookbooks (\url{http://isdc.unige.ch/?Support+documents}). For \xmm we
only use data from the EPIC-pn as the EPIC-MOS was switched off to allocate
maximum telemetry to the EPIC-pn (see also the paragraph on the Modified
Timing Mode below). The energy range used for the
\xmm analysis is 2.8 to 9.4\,keV.

Due to the not yet finalized calibration of the \xmm Modified Timing Mode we
divide our analysis in two parts. For the first part we combined the \inte
and \xte data to study the broadband continuum. The analysis of the
relativistically broadened Fe K$\alpha$ line was done independently using the
\xmm data only.

\section{Continuum}

To model the broadband continuum we used the hybrid thermal/non-thermal
Comptonization code \texttt{eqpair} by \cite{coppi92} which also includes
electron-positron pair production. In this model the temperature of the Comptonizing medium
is computed self-consistently by balancing Compton cooling with external
heating. The amount of heating is specified by the the ratio
$\ensuremath{\ell_\text{h}/\ell_\text{s}}$ of the compactness of the
Comptonizing medium and the seed photon distribution. We modeled the soft
emission by adding a \texttt{diskbb} component which provides the seed photons
for the Comptonization (therefore we set the temperature of the
\texttt{eqpair} seed photons equal to the temperature at the inner edge of the
disk). This continuum is partly reflected off the accretion disk (\xspec model
\texttt{reflect}) and modified at lower energies by interstellar absorption.

\begin{figure}
\centering
\includegraphics[width=\linewidth]{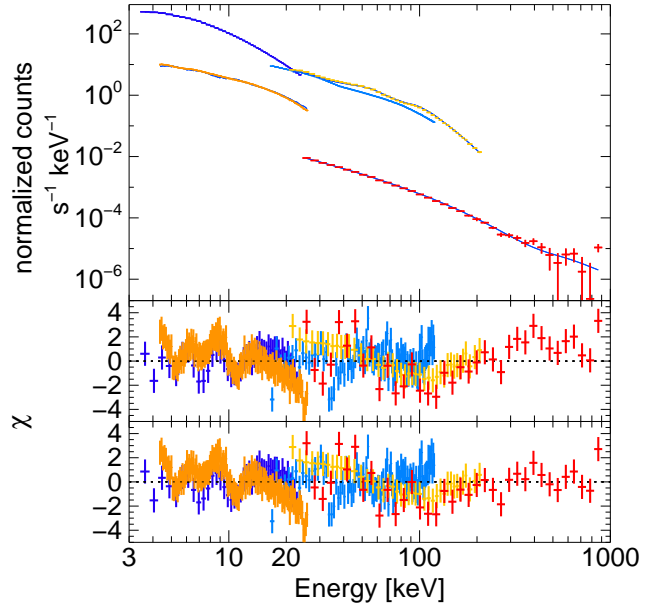}
\caption{Best fit to the time averaged spectrum using the
  \texttt{eqpair} model (\pca : 3--25\,\kev , dark blue; \hexte :
  15--120\,\kev ,
  light blue; \jemx : 4--26\,\kev , orange; \ibis : 20--250\,\kev , yellow; \spi :
  25--1000\,\kev , red). The upper panel shows the residuals to the pure
  thermal fit, the lower one those of the hybrid thermal/non-thermal model. \label{fig:eqpair}}
\end{figure}

We first performed pure thermal \eqpair fits to the four observations
independently \citep{fritz06}. In all cases they showed evidence of a spectral
hardening above $\approx$ 300\,\kev which could be an indicator for the
presence of a non-thermal electron component in the plasma. In order to get
better statistics for \spi in the crucial energy range we decided to sum up a
time averaged spectrum comprising all \inte and \xte observations although
\cyg was highly variable during the observations. For this time averaged
spectrum we first considered again a pure thermal plasma. We fixed
$\ell_\text{s}=10$ to force the seed photon spectrum to be dominated by disk
radiation and assumed a disk inclination of 45 deg
\citep[e.g.][]{gierlinski99}. As the inner radius of the disk could not be
constrained it was fixed to its default value
$R_\text{in}=10R_\text{g}$. Table~\ref{tab:eqpair} shows the best fit parameters. The
resulting disk parameters $kT_\text{in}=1.10^{+0.03}_{-0.02}$\,\kev and
norm=$45^{+7}_{-3}$ as well as the values for the compactness ratio
($\ensuremath{\ell_\text{h}/\ell_\text{s}} = 3.32^{+0.02}_{-0.01}$), the
optical depth ($\tau=1.23^{+0.01}_{-0.01}$), and the reflection covering
factor ($\Omega/2\pi=0.29^{+0.01}_{-0.01}$) are consistent with the results
obtained from analyzing the four observations independently. Also the Iron
line parameters match the values obtained before. The $\chi_\text{red}^{2}$
yields a value of 1.72 with 350\,dof. As shown in Fig.~\ref{fig:eqpair} a
deviation between the data points and this pure thermal model is still present
in the time averaged spectrum.

Therefore we added a non-thermal component in the \eqpair model by allowing
the parameter $\ensuremath{\ell_\text{nth}/\ell_\text{h}}$ to vary. We found
that 57\% of the power supplied to the electrons in the plasma is contained in
the non-thermal distribution of the electrons, which was chosen to be a
power law (Table~\ref{tab:eqpair}). This additional component improves
the $\chi_\text{red}^{2}$ to 1.62 (349\,dof). The dimensionless parameter
$\ensuremath{\ell_\text{h}/\ell_\text{s}}$ increases to a value of
$4.49^{+0.05}_{-0.03}$, $\tau=1.47^{+0.01}_{-0.01}$ and
$\Omega/2\pi=0.33^{+0.01}_{-0.01}$ are also slightly higher than in the pure
thermal model. The disk parameters ($kT_\text{in}=1.09^{+0.07}_{-0.03}$\,\kev and
norm=$48^{+9}_{-3}$) as well as the values found for the Fe K$\alpha$ line do
not change significantly with respect to the thermal model.

\begin{table}
\caption{Best fit parameters for the \eqpair model. The iron line is modeled
  as a simple Gaussian here.}
\label{tab:eqpair}
\begin{tabular}{l@{\hspace*{1mm}}c@{\hspace*{1mm}}c@{\hspace*{1mm}}} \hline\hline\\
                                         & \eqpair (th.) &\eqpair (th./nth.) \\[1ex] \hline\\
$N_\text{H}$ [$10^{22}\text{cm}^{-2}$]    & $0^{+0.22}_{-0}$          & $0^{+0.16}_{-0}$       \\[1.3ex]
$kT_\text{in}$ [\kev]                      & $1.10^{+0.03}_{-0.02}$    & $1.09^{+0.07}_{-0.03}$ \\[1.3ex]
norm                                      & $45^{+7}_{-3}$            & $48^{+9}_{-3}$        \\[1.3ex]
$E_{\text{K}\alpha}$ [\kev]               & $6.32^{+0.12}_{-0.12}$    & $6.32^{+0.05}_{-0.11}$ \\[1.3ex]
$\sigma_{\text{K}\alpha}$ [\kev]          & $0.77^{+0.13}_{-0.06}$    & $0.72^{+0.12}_{-0.06}$ \\[1.3ex]
$\ensuremath{\ell_\text{h}/\ell_\text{s}}$& $3.32^{+0.02}_{-0.01}$    & $4.49^{+0.05}_{-0.03}$ \\[1.3ex]
$\ensuremath{\ell_\text{nth}/\ell_\text{h}}$& --                      & $0.57^{+0.02}_{-0.03}$ \\[1.3ex]
$\tau$                                    & $1.23^{+0.01}_{-0.01}$    & $1.47^{+0.01}_{-0.01}$ \\[1.3ex]
$\Omega/2\pi$                             & $0.29^{+0.01}_{-0.01}$    & $0.33^{+0.01}_{-0.01}$ \\[1.3ex]
$\xi$                                     & $2^{+7}_{-2}$             & $0^{+19}_{-0}$          \\[1.3ex]
                                          &                           &                        \\[1ex]
$\chi_\text{red}^{2}$ / dof               & 1.72\,/\,350              & 1.62\,/\,349           \\[1.5ex] \hline
\end{tabular}
\end{table}

\section{Iron line}

\subsection{Modified Timing Mode}
Due to telemetry restrictions it is not always trivial to study bright sources
with the maximum possible time resolution in combination with a satisfying
signal to noise ratio using the standard modes of \xmm. In the EPIC-pn burst
mode (which is foreseen for very bright sources) only 3\% of all detected
photons are transmitted resulting in a enormous reduction of the signal to
noise ratio. We therefore used an alternative approach which includes
switching off the EPIC-MOS camera and running the EPIC-pn in a Modified Timing
Mode \citep{kendziorra04}. In this mode the lower energy threshold was
increased to 2.8\,\kev (the standard value is 200\,eV). This implies that a
re-calibration of the instrument is required because the combination of split
events is not done on-board but during the first step of the EPIC-pn data
analysis. Due to the increased lower threshold a large fraction of the split
partners is not transmitted and therefore the spectrum appears to be
softer. By comparison with former Timing Mode observations we built a new
detector response matrix \citep{wilms06b}. However, there are still some effects not included
yet, for example the improvement of the Charge Transfer Efficiency of the
detector due to the high count rates. The special calibration for the Modified
Timing Mode will be made public through the \xmm SOC as soon as full
confidence on the calibration has been reached.

\subsection{Relativistically broadened Iron lines}
The Fe K$\alpha$ line observed in BH candidates is intrinsically
narrow with a rest frame energy of 6.4--6.97\,keV depending on the ionization
state. It originates from material which is just a few gravitational radii
away from the BH and therefore it is broadened by gravitational
redshift effects as well as by Doppler shifts. The exact shape of the line
depends on the accretion geometry, namely on the Fe K$\alpha$ emissivity of
the disk, the angular momentum of the BH and the observer's viewing
angle. Relativistically broadened Iron lines are therefore an important
observational tool for the understanding of the accretion geometry. For more
information on relativistically broadened Iron lines see for example \citep{reynolds03,fabian06}.

\subsection{Results}
\begin{figure}
\centering
 \includegraphics[width=\linewidth]{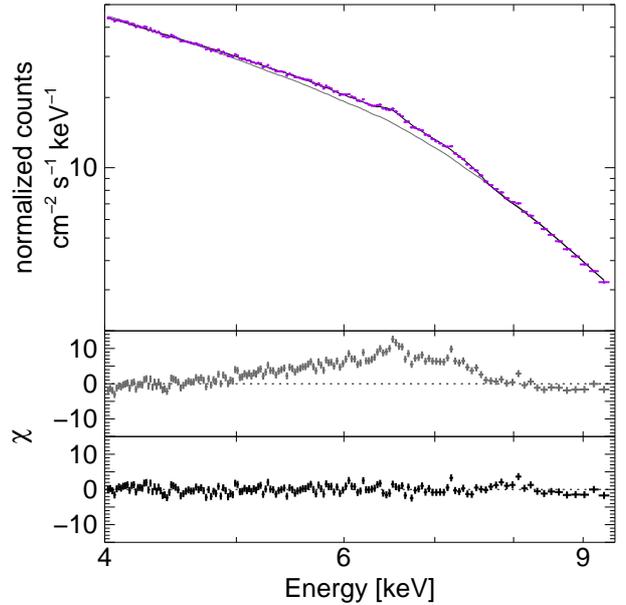}
\caption{Fit to the obs2 \xmm data. \texttt{Upper panel:} Power-law fit to the data
  outside the 5--8\,keV band. There are strong residuals remaining in the Fe
  K$\alpha$ region. \texttt{Lower panel:} Fit to the whole band using a
  power-law and a narrow line and a relativistic line.\label{fig:xmm}}
\end{figure}

Due to the increased soft X-ray emission during the transitional state there
is an increased pile-up in the center of the point spread function. We
therefore excluded data from the innermost 3 CCD columns from our analysis,
however the signal to noise ratio of the data is still sufficient for spectral
analysis as could be seen in Fig.~\ref{fig:xmm}.

A power-law fit to the \xmm data outside the 5--8\,keV band reveals strong
residuals in the Fe K$\alpha$ region ($\chi^{2}_\text{red}=21.4$). Modeling
these residuals by a narrow ($\sigma=80\pm35$\,eV) line at
$E=6.52\pm0.02$\,keV with an equivalent width of 14\,eV and a relativistic
Kerr line at $E=6.76\pm0.1$\,keV (equivalent width 400\,eV) improves the
$\chi^{2}_\text{red}$ dramatically to 1.3. The emissivity of the relativistic
line is found to be $\propto r^{-4.3\pm0.1}$. These parameters are similar
to earlier \textsl{Chandra} intermediate state observations by Miller et
al. \citep{miller02}. Note however that the calibration of the timing mode is
not yet fully completed so that the energy values obtained from our analysis
could be slightly too high.

\section{Summary and Outlook}
The 3\,\kev--1\,MeV broadband spectrum of \cyg can be well described by the \texttt{eqpair}
model. The parameters we obtain from our analysis are consistent with previous
results e.g. from the \xte monitoring campaign
\citep{wilms06a,fritz06,pottschmidt06}. Furthermore our data show evidence for the presence
of a non-thermal component in the distribution of the electrons with 57\% of
the power supplied to the electrons going into the non-thermal acceleration.

The simultaneous \xmm observation confirms the presence of a relativistic line
\citep{miller02} during the intermediate state of \cyg and the parameters
obtained for the broad and narrow components lead to the conclusion that the
accretion disk seems to extend down to the innermost stable orbit at $1.235\,R_\text{g}$.

As soon as the calibration of the Modified Timing Mode is completed we will
combine all three instruments to model the whole broadband spectrum from
2.8\,keV up to 1\,MeV.

\bibliographystyle{plain}
\bibliography{mnemonic,jw_abbrv,bhc}

\end{document}